# ABSTRACT

# TREND WITHOUT HICCUPS - A KALMAN FILTER APPROACH

By

ERIC BENHAMOU
DATE: April 2016

Have you ever felt miserable because of a sudden whipsaw in the price that triggered an unfortunate trade? In an attempt to remove this noise, technical analysts have used various types of moving averages (simple, exponential, adaptive one or using Nyquist criterion). These tools may have performed decently but we show in this paper that this can be improved dramatically thanks to the optimal filtering theory of Kalman filters (KF). We explain the basic concepts of KF and its optimum criterion. We provide a pseudo code for this new technical indicator that demystifies its complexity. We show that this new smoothing device can be used to better forecast price moves as lag is reduced. We provide 4 Kalman filter models and their performance on the SP500 mini-future contract. Results are quite illustrative of the efficiency of KF models with better net performance achieved by the KF model combining smoothing and extremum position.

# ACKNOWLEDGEMENTS

The author would like to acknowledge support from Thomson Reuters. Data are from Thomson Reuters Eikon while screenshots and source codes are done with TR Eikon Trading Robot. The author also thanks Denis Dollfus for fruitful conversations.

# INTRODUCTION

## 1 INTRODUCTION

Have you ever felt angry because a sudden price whipsaw triggered an unfortunate signal and a resulting bad trade? Prices have inherent blips and jerks that are not easy to control. Moreover, prices are inputs for technical analysis indicators. This can result in corrupted or non-efficient indicators. In an ideal world, one would like prices heading to a clear direction. Remember the old adage: "trade with the trend". But in real life, prices hiccups create noise and perturb the signal.

A first attempt to remove these yanks and jolts is to smoothen prices with moving averages. However, moving averages suffer from two flaws: lags and no dynamics. The first drawback, delay in moving average response, is widely known as moving averages used past data. Adaptations to moving averages have been suggested (exponential, adaptive, zero lag or Nyquist criterion based moving averages). Durchner in **[Dürschner-2012]** suggested the use of Nyquist criterion to create moving average 3.0 with no lag.

This is intellectually very enticing as the lag is completely removed. This improves moving averages from Patrick Mulloy **[Mulloy-1994]** with zero lag or the attempts by John Ehlers to provide sophisticated moving averages (**[Ehlers-2001a]** or **[Ehlers-2001b]**). But this does not address the second problem of capturing price dynamics. What we mean by price dynamics is the price movement. If we can identify that prices are moving upwards (respectively downwards), then a good guess for next price observation should be higher (respectively lower) than the current price. Let us pause for a moment and imagine instead of prices, we were looking at car position using a GPS. We measure the car position with a GPS but with some noise as the signal is not perfectly accurate. Could we capture the car dynamics to compute the best guess at next time step and hence reduce noise in car position? The answer is yes! And guess what, this is what your car GPS is doing. This theory simply explained is referred to as Kalman filter (from its inventor **[Kalman-1960]**) shortened to KF in this paper. It was created for the spatial industry to remove noise and capture shuttle movements. In a scientific way, the Kalman filter is an efficient recursive filter that estimates the state of a dynamic system from a series of incomplete and noisy measurements to estimate the best forecast according to an assumed distribution. In the original paper, Kalman assumes a Gaussian distribution of noise but extended version can now cope with more advanced distribution (see **[Kalman-Filter-Wikipedia]**). In this article, we first revisit moving averages and then present different Kalman filter models and their implementation to create trading strategies. We then provide performance results for our 4 KF models on one year of data of the E-mini-SP continuation future.

**1.1 Motivation for smoothing**

Smoothing prices is natural. The basic idea is to remove noise from prices to better identify important patterns or trends. Remember, when we trade, we want the big picture. So smoothing enables us removing bumps, bangs, bounces, and shocks and getting an average clean signal. If we believe that prices do not follow a random walk model, the smoothened signal provides us a clear directional signal.

**1.2 Impact for trading strategies**

Conversely, if we do not smoothen prices, we could act on tugs, wrenches, or snatches that are against the trend and result in bad trades. Smoothing is the right way! But we need to be careful. If we smoothen with lag (one of the major drawbacks of moving averages), we act with delay and enter trades too late, potentially facing reverse direction markets. In an ideal world, we would like the smoothing technique to have zero lag and to provide a fist move advantage.

<div style="text-align:center">**MATERIAL AND METHODS**</div>

## 2 REVIEW OF MOVING AVERAGES

### 2.1 The usual moving averages

The usual way to remove noise in prices is with moving averages. Let us denote weights by $w_i$ for the time $T_i$, where $i$ goes between 0 to $N$. Then, the moving average is given by

$$MA(w_0, \ldots, w_N) = \sum_{i=0}^{N} w_i P_i \qquad \text{(EQ 2.1)}$$

Whose lag is

$$LAG\ MA(w_0, \ldots, w_N) = \frac{\sum_{i=0}^{N} w_i T_i}{\sum_{i=0}^{N} w_i} \qquad \text{(EQ 2.2)}$$

If we do a moving average of a moving average, equation (2.1) becomes

$$MA\odot MA(w_0, \ldots, w_N) = \sum_{i=0}^{N} w_i \left( \sum_{j=0}^{N} w_j P_{i+j} \right) \qquad \text{(EQ 2.3)}$$

And the corresponding lag is

$$LAG\ MA\odot MA(w_0, \ldots, w_N) = \frac{\sum_{i=0, j=0}^{N} w_i w_j T_{i+j}}{\sum_{i=0, j=0}^{N} w_i w_j} \qquad \text{(EQ 2.4)}$$

We can easily derive similar formula for a recursive moving average at the order kth:

$$MA\odot^k(w_0, \ldots, w_N) = \sum_{i_1=0, \ldots, i_k=0}^{N} w_{i_1} \ldots w_{i_k} P_{i_1 + \cdots + i_k} \qquad \text{(EQ 2.5)}$$

The resulting lag is

$$LAG\ MA\odot^k(w_0, \ldots, w_N) = \frac{\sum_{i_1=0, \ldots, i_k=0}^{N} w_{i_1} \ldots w_{i_k} T_{i_1 + \cdots + i_k}}{\sum_{i_1=0, \ldots, i_k=0}^{N} w_{i_1} \ldots w_{i_k}} \qquad \text{(EQ 2.6)}$$

### 2.2 Explicit Lag Computation

Prices are sampled with equidistant time steps $T_i = \frac{i}{N}T$. Formulae (EQ.2.6) can be easily computed in terms of its first order value as follows:

$$LAG\ MA\odot^k(w_0, \ldots, w_N) = k\ LAG\ MA(w_0, \ldots, w_N) \quad (EQ\ 2.7)$$

(See **Proof A.1:**)

Furthermore, if we combine recursive moving averages, it is easy to find back the results of Mulloy. In the case of a moving average of moving average, the only possible choice with zero lag whose coefficient sum is equal to 1 is the double moving average:

$$DEMA = 2MA\text{-}MA\odot^2 \quad (EQ\ 2.8)$$

(See **Proof A.2**)

And for the triple moving average (if we impose the additional constraint that the $3^{rd}$ order recursive moving average coefficient is 1), we have

$$TEMA = 3MA\text{-}3.MA\odot^2 + MA\odot^3 \quad (EQ\ 2.9)$$

(See **Proof A.3**)

## 3  INTRODUCTION TO KALMAN FILTER

### 3.1 Basic concepts

Kalman filter is a recursive algorithm that has been invented in the 1960s to track a moving target, remove any noisy measurements of its position and predict its future position. In finance, KF has been used by the asset management industry for various purposes. KF is an optimal choice in many cases and do at least better than a moving average smoothing. Dao et al **[Bruder-Dao-Richard-Roncalli-2011]** and **[Dao-2011]** showed that for price following random walk with noise, KF is equivalent to the optimal exponential moving average with parameter equal to Kalman gain. However, for more sophisticated dynamics, like a linear Gaussian model, KF is the optimal choice and the most efficient computational solution for finding the model parameters.

KF has also been used over the last decade by different authors. Martinelli and Rhoads in **[Martinelli-2006]** and **[Martinelli-Rhoads-2010]** used Kalman filter to find optimal guess for trading strategies on stocks. Haleh-et-al in **[Haleh-et-al-2011]** used Extended Kalman filter for forecasting stock prices, combining technical and fundamental data. They showed it outperformed regression and neural networks. Ernie Chan in **[Chan-2013]** suggested using KF for pair correlation trading, while Cazalet and Zheng in **[Cazalet-Zheng-2014]** used KF for hedge fund replication.

In a general way, Kalman filter consider a linear dynamic system given by

$$X_{t+1} = \phi X_t + c_t + w_t \tag{EQ 3.1}$$

$$Y_t = H X_t + d_t + v_t \tag{EQ 3.2}$$

Where $\phi$ is the state transition matrix, $H$ the measurement matrix, $w_t$ the model noise, $X_t$ the state vector, $Y_t$ the measurement vector, $v_t$ the measurement noise, $w_t$ and $v_t$ the independent white noises with zero mean and their variance matrices given by $Q$ and $R$ respectively. $c_t$, respectively $d_t$, is the drift of the state vector, respectively the measurement vector. The corresponding Kalman filter is:

Prediction step: $\quad X_{t+1|t} = \phi X_{t|t} + c_t \tag{EQ.3.3}$

With $\quad P_{t+1|t} = \phi P_{t|t} \phi^T + Q \tag{EQ.3.4}$

Correction step: $\quad X_{t+1} = X_{t+1|t} + K_{t+1}(Y_{t+1} - Y_{t+1|t}) \tag{EQ.3.5}$

With $\quad Y_{t+1|t} = H X_t + d_t$

With Kalman gain $\quad K_{t+1} = P_{t+1|t} H^T [H P_{t+1|t} H^T + R]^{-1} \tag{EQ.3.6}$

With $\quad P_{t+1|t+1} = [I - K_{t+1} H] P_{t+1|t} \tag{EQ.3.7}$

KF works in a two-step process (prediction and correction steps). The algorithm is recursive, can run in real time, using only the present input measurements, the previously calculated state and its uncertainty matrix.

Obviously, one needs to specify the state and measurement vector. A logic choice is to use a physical system with concepts similar to speed and acceleration:

$$x_{t+1} = x_t + \dot{x}_t \delta t + 1/2\, a_t \delta t^2 \qquad (EQ.3.8)$$

$$\dot{x}_{t+1} = \dot{x}_t + a_t \delta t \qquad (EQ.3.9)$$

$$y_t = x_t + v_t \qquad (EQ.3.10)$$

Where $x_t$ and $\dot{x}_t$ are price and rate of change of stock price at time $t$ (similar to position and speed). $a_t$ can be seen as the acceleration of price at time $t$. It is considered to be a model noise. $T$ is the sampling period, $y_t$ the measurement, $v_t$ the measurement noise. This can be analyzed as a KF system with

$$X_t = \begin{bmatrix} x_t \\ \dot{x}_t \end{bmatrix}, \phi = \begin{bmatrix} 1 & \delta t \\ 0 & 1 \end{bmatrix}, w_t = \begin{bmatrix} \frac{1}{2}\delta t^2 \\ \delta t \end{bmatrix} a_t, H = \begin{bmatrix} 1 & 0 \end{bmatrix}, c_t = 0 \qquad (EQ.3.11)$$

This is named **model One**. This model has the advantage to take into account a certain dynamic compared to the simple Random Walk model that is often used in the KF literature where there is no speed term. In our model One, The speed is initially estimated as the difference between two consecutive prices. The parameters to estimate are the following (4 in total)

$$Q = \begin{bmatrix} p_1 p_1 & p_2 p_1 \\ p_1 p_2 & p_2 p_2 \end{bmatrix}, R = [p_3], P_{t=0} = \begin{bmatrix} p_4 & 0 \\ 0 & p_4 \end{bmatrix} \qquad (EQ.3.12)$$

It is interesting to note that this model is very closed to a local linear trend model. Indeed, the local linear trend model writes as

$$x_{t+1} = x_t + \beta_t + w_{1,t} \qquad (EQ.3.13)$$

$$\beta_{t+1} = \beta_t + w_{2,t} \qquad (EQ.3.14)$$

$$y_t = x_t + v_t \tag{EQ.3.15}$$

We can notice that in this specific case, the KF parameters are the following:

$$X_t = \begin{bmatrix} x_t \\ \beta_t \end{bmatrix}, \phi = \begin{bmatrix} 1 & 1 \\ 0 & 1 \end{bmatrix}, w_t = \begin{bmatrix} w_{1,t} \\ w_{2,t} \end{bmatrix}, H = \begin{bmatrix} 1 & 0 \end{bmatrix}, c_t = 0 \tag{EQ.3.16}$$

The parameters to estimate are the following (5 in total)

$$Q = \begin{bmatrix} p_1 p_1 & p_2 p_1 \\ p_1 p_2 & p_2 p_2 \end{bmatrix}, R = [p_3], P_{t=0} = \begin{bmatrix} p_4 & 0 \\ 0 & p_5 \end{bmatrix} \tag{EQ.3.17}$$

This model has almost the same parameters as model one. This is named **model Two**. Comparing equation 3.12 and 3.17, we know that model One and Two should have very similar behavior.

We can create a more general two factors model with contribution to price split between a short term $x_t^1$ and a long term $x_t^2$. This leads to:

$$x_{t+1}^1 = a_{11} x_t^1 + a_{12} x_t^2 + w_{1,t} \tag{EQ.3.18}$$

$$x_{t+1}^2 = a_{22} x_t^2 + w_{2,t} \tag{EQ.3.19}$$

$$y_t = h_1 x_t^1 + h_2 x_t^2 + v_t \tag{EQ.3.20}$$

In this specific model, we have the following parameters

$$X_t = \begin{bmatrix} x_t^1 \\ x_t^2 \end{bmatrix}, \phi = \begin{bmatrix} a_{11} & a_{12} \\ 0 & a_{22} \end{bmatrix}, w_t = \begin{bmatrix} w_{1,t} \\ w_{2,t} \end{bmatrix}, H = \begin{bmatrix} h_1 \\ h_2 \end{bmatrix}, c_t = 0 \tag{EQ.3.21}$$

We call this **model Three**. Because of its generality, this model encompasses models 1, 2.

The parameters to estimate are the following (10 in total)

$$\phi = \begin{bmatrix} p_1 & p_2 \\ 0 & p_3 \end{bmatrix}, H = \begin{bmatrix} p_4 \\ p_5 \end{bmatrix}, Q = \begin{bmatrix} p_6 p_6 & p_7 p_6 \\ p_6 p_7 & p_7 p_7 \end{bmatrix}, R = [p_8], P_{t=0} = \begin{bmatrix} p_9 & 0 \\ 0 & p_{10} \end{bmatrix} \tag{EQ.3.22}$$

The last model we use is a model inspired by a combination of oscillators and the previous model. In this model, we use the price position with respect to its extremums as in the fast stochastic oscillator. We denote the variable $K_t$ over a d period given by

$$K_t^d = \frac{\text{Current Close} - \text{Lowest Low(d)}}{\text{Highest High(d)} - \text{Lowest Low(d)}} \times 100 \tag{EQ.3.23}$$

We denote by $L_t^d = Lowest\ Low(d)$ and $H_t^d = Highest\ High(d)$ the lowest low and highest high over d period. We use in our example a 14 days period. As in model 3, we also split the contribution of the price due to short term $x_t^1$ and long term $x_t^2$. This leads to:

$$x_{t+1}^1 = a_{11}x_t^1 + a_{12}x_t^2 + (M_1 - N_1 K_t^d) + w_{1,t} \quad (EQ.3.24)$$

$$x_{t+1}^2 = a_{22}x_t^2 + (M_2 - N_2 K_t^d) + w_{2,t} \quad (EQ.3.25)$$

With
$$K_t^d = \frac{h_0 x_t - L_t^d}{H_t^d - L_t^d} \quad (EQ.3.26)$$

$$y_t = h_1 x_t^1 + h_2 x_t^2 + v_t \quad (EQ.3.27)$$

In this specific model, we have the following parameters

$$X_t = \begin{bmatrix} x_t^1 \\ x_t^2 \end{bmatrix}, \Phi = \begin{bmatrix} a_{11} & a_{12} \\ 0 & a_{22} \end{bmatrix}, w_t = \begin{bmatrix} w_{1,t} \\ w_{2,t} \end{bmatrix}, H = \begin{bmatrix} h_1 \\ h_2 \end{bmatrix}, c_t = \begin{pmatrix} M_1 - N_1 K_t^d \\ M_2 - N_2 K_t^d \end{pmatrix} \quad (EQ.3.28)$$

We call this **model Four**. Because of its generality, this model encompasses models 1, 2 and 3. It capture short and long term effect as well as position with regards to extrema like what oscillators do. This is by far the most realistic model. Short term factor $x_t^1$ models extreme market reactions that last for a few days. Long term factor $x_t^2$ is only influenced by itself and not by the short term $x_t^1$. The parameters to estimate are the following (15 in total) (the same set as model 3 and 5 additional parameters)

$$\Phi = \begin{bmatrix} p_1 & p_2 \\ 0 & p_3 \end{bmatrix}, H = \begin{bmatrix} p_4 \\ p_5 \end{bmatrix}, Q = \begin{bmatrix} p_6 p_6 & p_7 p_6 \\ p_6 p_7 & p_7 p_7 \end{bmatrix}. R = [p_8], P_{t=0} = \begin{bmatrix} p_9 & 0 \\ 0 & p_{10} \end{bmatrix} \quad (EQ.3.29)$$

$$c_t = \begin{pmatrix} p_{11} - p_{12} K_t^d \\ p_{13} - p_{14} K_t^d \end{pmatrix}, p_{15} = d \quad (EQ.3.30)$$

### 3.2 Pseudo code

```
/// Initialization phases: parameters contains
///      - initial value for model state + measurement of model
///      - measurement of state and model variance
Kalman2D k = new Kalman2D(parameters);
k.Setup( parameters );
int length = timeSeries.Length;

Point2D[] kalmanResult = new Point2D[length];
```

```csharp
/// the loop to update in real time
for( int i = 0; i<length; ++i )
{
        if( i<Period )
        {
                k.Predict();
                k.Update(timeSeries[i]);

                kalmanResult.Set(0, timeSeries[i]);
                kalmanResult.Set(1, timeSeries[i]);
        }
        else
        {
                k.Predict();
                kalmanResult.Set(0, k.X.Get(0,0) );
                k.Update(timeSeries[i]);
                kalmanResult.Set(1, k.X.Get(0,0) );
        }
}
```

Where the pseudo code for Predict and Update value is given as follows:

```csharp
// <summary>
// Predict the state
// </summary>
public void Predict()
{
   // Predict to now, then update.
   // Predict:
   //   X = Phi*X + C
   //   P = Phi*P*Phi^T + Q
   m_x = Phi*m_x+C;
   m_p = Phi*P*Transpose(Phi) + Q;
}

// <summary>
// Update the state thanks to the realized measurement Y
// </summary>
public void Update( Point1D Y )
{
   // Update:
   //   I = Y-(HX+D) Called the innovation= measurement – state transformed by H.
   m_I= Y(t)-(H*m_X+D);
   //   S = H*P*H^T + R   S= Residual covariance = covariance transformed by H + R
   m_S= H*m_P*Transpose(T) + R;
   //   K = P * H^T *S^-1   K = Kalman gain = variance / residual covariance.
   m_K= m_P * Transpose(H) * Inverse(S);
   //   X = X + K*I    Update with gain the new measurement
   m_X += m_K*Y;
   //   P = (I – K * H) * P  Update covariance to this time.
   m_P= (I – K * H) * P;
}
```

# 4 TRADING STRATEGIES WITH KALMAN FILTER

## 4.1 Basic concepts

KF model enables us various things:

- It smoothens any data. Hence the data produced by the KF can be used instead of prices to remove any spike. This opens multiple options as these inputs can be used in cross over moving averages strategies, MACD indicator, oscillators and combination of these. We do not explore this as the paper goal is to study the predictive power of KF models.

-it can be used as a predictive tool to help deciding when entering long or short strategies. We compare the prediction with the current. This is precisely the subject of this paper.

## 4.2 Pseudo code

```
/// <summary>
/// Called on each new bar event
/// </summary>
protected override void OnNewBar()
{
if (KalmanFilter(Param1,..,ParamN).Predict[0] > Close[1]+Offset)
        EnterLong();
else if (KalmanFilter(Param1,..,ParamN).Predict[0] < Close [1]-Offset)
        EnterShort();}
}
```

# RESULTS

## 5 NUMERICAL RESULTS

### 5.1 Description of the sample set

To test the efficiency of KF models 1, 2, 3 and 4, we use the E-mini-S&P-500 continuation Future, whose RIC is **Esc1**. We use the Eikon App "Trading Robot" that has been developed by the author. We look at daily data between 28 Feb 2015 and 28 Feb 2016.

### 5.2 Comparison of Kalman filters with standard technical indicators

We provide graphics of various indicators to measure how KFs best fit price information. We display:

- some standard technical analysis indicators:

    - Moving averages with lag: standard and exponential moving average with 12 days period

    - Moving averages with zero lag: double exponential moving average with 12 days period as (EQ.2.9) and triple exponential moving average with 12 days period as (EQ.2.10)

- the different KF indicators, KF model 1, 2, 3 and 4.

In Figure 1, we see that the KF model 1 sticks much better to price data than any of the two moving averages. This is normal as KF model has 0 to 1 period lag. We do not show in this graphic the other KF models as they would be barely distinguishable. In Figure 2 and Figure 3, we compare KF model with zero-lag moving averages like DEMA or TEMA. We emphasize area of difference with orange circles and see that KF models stick much better to price data. In Figure 4, we compare the different KF models and see that KF model 1&2 are similar while models 3&4 are also similar, with an advantage to the latter ones.

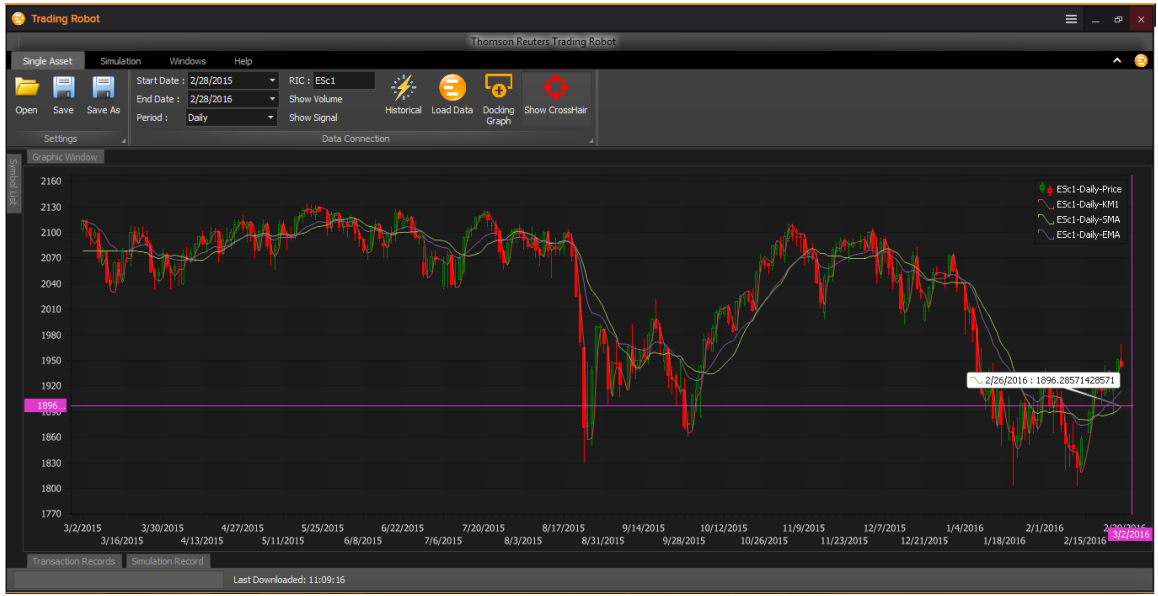

*Figure 1: Comparison of Kalman filter with classical moving averages.*
*The red line representing the KF model 1 sticks much better to the price data than any of the two moving averages (standard and exponential ones with both 12 periods).*

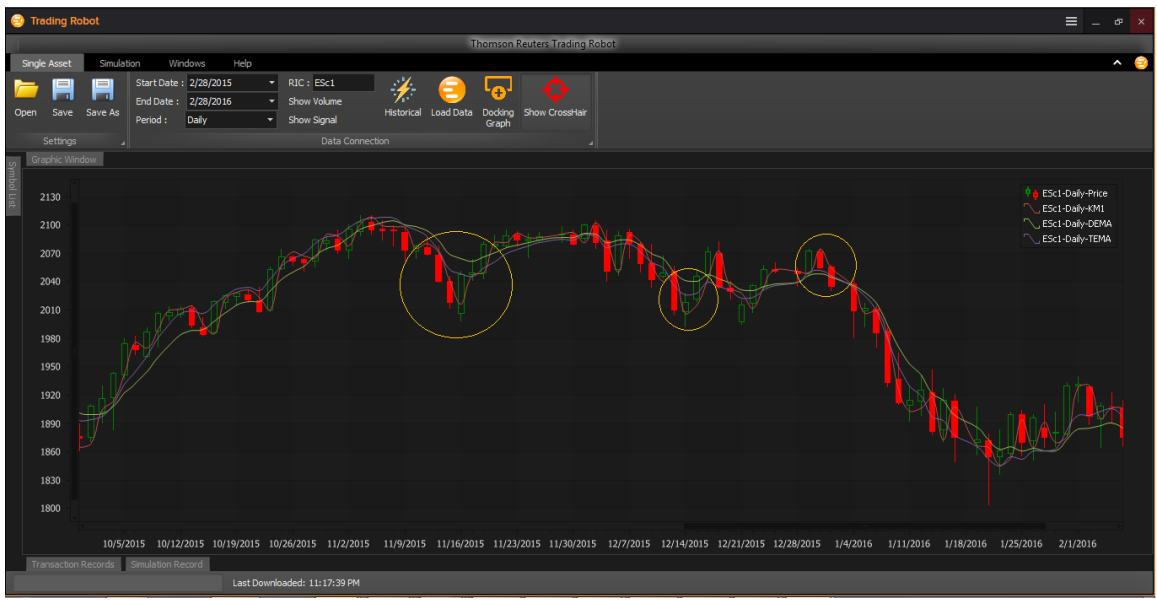

*Figure 2: Comparison of Kalman filter with double and triple exponential moving averages*
*Red lines representing the KF model 1 stick much better to the price data than DEMA or TEMA as displayed in orange circles.*

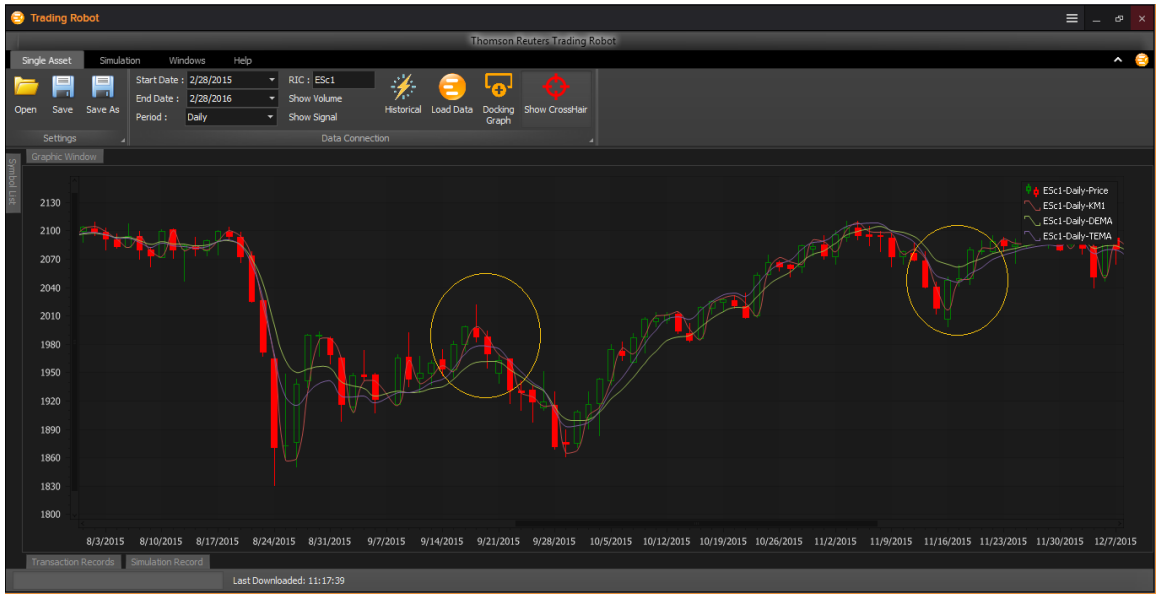

*Figure 3: Zoom on differences between Kalman filter and zero lag moving averages.*
*KF model 1 reacts faster to price changes as emphasized by orange circles.*

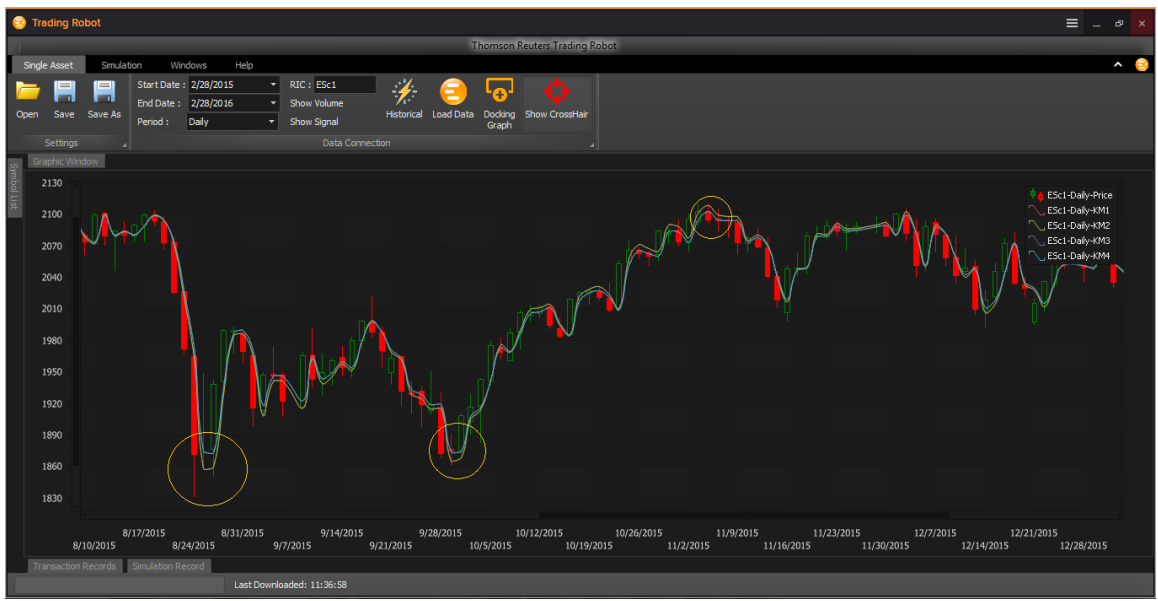

*Figure 4: Comparison between the different Kalman filters.*
*Within the KF model family, model 3 and 4 are even better than model 1 and 2. Model 1 and 2 (respectively model 3 and 4) have similar behaviors.*

**5.3 Kalman filter trading strategies performance**

We look at the same one-year of data and compute the optimal parameters for the 4 KF models. For each model, we use no leverage, trade only one future contract regardless of the current trading account. We also assume a 4 dollar round trip commission, which is the observed price at retailed brokers like Interactive-Brokers. For large trader with more than 20,000 contracts per months and CME membership, round trip commission lowers to 1.4 USD.

Table 1 shows that the best model is model 4 with an annual net profit of 39.5 k USD, followed by model 3 with 29 k USD and the last two being model 2 and 1 with net profit of 22 and 19 k USD.

We can make various remarks:

- The final model ranking makes senses as model 4 is richer than 3 that is itself richer than 2 that is richer than 1.
- The best model, KF 4, provides a nice net profit, 39 k with a max drawdown of -2,600, hence representing a ratio of net profit over drawdown (also called recovery ratio) of 15. This is excellent!
- E-mini-S&P daily margin is about 5 to 6 k USD, hence 40k USD net profit is an amazing statistic. In addition, model 4 incurs only positive monthly PnL (Figure 5).
- KF model 3 has a nice and steady cumulative profit curve (Figure 6), while model 4 outperforms because it captures a few large additional trades (Figure 5 and Figure 9).
- KF model 1 & 2 are Kalman filter models already explored in literature. We find some negative monthly PnL and large drawdown (see Figure 7 and Figure 8). This is a known feature as these models have a poor dynamic. This may explain why these standard KF models have been disregarded.
- The difference between KF model 3 and 4 is the oscillator factor. This confirms the well-known fact that oscillators capture others features than trending indicators and catch any mean reverting market (in trading range environment). The combination of trend following factors (like in model 3) with the new extra term inspired from oscillators yields a powerful model called 4. We can notice that parameter 14, $N_2$, is null. It indicates that the oscillator factor plays a role only on short term factor. This can be interpreted as an empirical evidence that range trading has only influence on short term while trend dominates in long term.
- The parameters 11 and 13 in KF model 4 represent the neutrality level at which the oscillator factors changes from bullish to bearish. It is amazing that its optimal value turns out to be 50%, which is also a well-known feature of oscillators where the level of neutrality is 50%

We provide optimal parameters in Table 2. We also provide various statistics for KF model 4, 3, 2 and 1 (starting with the best model and going to the worst) in Table 3, Table 4, Table 5, Table 6 and the list of all trades in Table 7

We provide in Figure 5, Figure 6,Figure 7, Figure 8 the cumulative profit and loss curve for trading strategy of model 4, 3, 2 and 1, starting by the best one. Figure 9 zooms on the period where model 4 locks in large profit due to accurate prediction of turning points.

| Model | Net Profit | Gross Profit | Gross Loss | Drawdown | Trades | Commission | Recovery ratio | Sharpe Ratio |
|---|---|---|---|---|---|---|---|---|
| Kalman Filter 1 | 18,755 | 39,151 | -20,396 | -7,348 | 55 | 220 | 2.55 | 0.72 |
| Kalman Filter 2 | 22,380 | 40,747 | -18,367 | -7,348 | 55 | 220 | 3.05 | 0.76 |
| Kalman Filter 3 | 29,022 | 47,548 | -18,526 | -3,800 | 57 | 228 | 7.64 | 1.22 |
| Kalman Filter 4 | 39,558 | 50,243 | -10,685 | -2,600 | 48 | 192 | 15.21 | 0.73 |

*Table 1: Trading performance of Kalman filter model 1, 2, 3 and 4.*

| Model | Kalman Filter 1 | Kalman Filter 2 | Kalman Filter 3 | Kalman Filter 4 |
|---|---|---|---|---|
| Parameter 1 | 5.00 | 5.00 | 1.00 | 1.00 |
| Parameter 2 | 5.00 | 5.00 | 0.40 | 0.40 |
| Parameter 3 | 45.00 | 41.00 | 1.20 | 1.20 |
| Parameter 4 | 10.00 | 1.00 | 1.00 | 1.00 |
| Parameter 5 |  | 1.00 | 1.00 | 1.00 |
| Parameter 6 |  |  | 0.80 | 0.80 |
| Parameter 7 |  |  | 0.40 | 0.40 |
| Parameter 8 |  |  | 0.70 | 0.70 |
| Parameter 9 |  |  | 1.00 | 1.00 |
| Parameter 10 |  |  | 0.40 | 0.40 |
| Parameter 11 |  |  |  | 0.50 |
| Parameter 12 |  |  |  | 0.90 |
| Parameter 13 |  |  |  | 0.50 |
| Parameter 14 |  |  |  | - |
| Parameter 15 |  |  |  | 5.00 |

*Table 2: Model parameters for Kalman filter model 1, 2, 3 and 4*

| Kalman Filer Model 4 | | | |
|---|---:|---:|---:|
| Field | All | Long | Short |
| **Net Profit (A+B)** | **39,558** | **17,279** | **22,279** |
| Gross Profit (A) | 50,243 | 20,957 | 29,286 |
| Gross Loss (B) | (10,685) | (3,678) | (7,007) |
| Total Commission | 192 | 96 | 96 |
| Drawdown | (2,600) | (2,137) | (2,520) |
| Sharpe Ratio | 0.73 | 0.84 | 0.55 |
| Profit Factor (A/B) | 4.70 | 5.70 | 4.18 |
| | | | |
| **Number of Trades** | **48** | **24** | **24** |
| Winning Trades | 30 | 17 | 13 |
| **Average Trade Profit** | **824** | **720** | **928** |
| Average Winning Trade | 1,675 | 1,233 | 2,253 |
| Largest Winning Trade | 11,309 | 5,434 | 11,309 |
| Max. conseq. Winners | 6 | 5 | 3 |
| | | | |
| Losing Trades | 18 | 7 | 11 |
| Average Losing Trade | (594) | (525) | (637) |
| Largest Losing Trade | (1,729) | (1,729) | (1,267) |
| Max. conseq. Losers | 4 | 2 | 3 |
| | | | |
| Ratio avg. Win / avg. Loss | 2.82 | 2.35 | 3.54 |
| Winning/Total | 0.63 | 0.71 | 0.54 |
| | | | |
| Avg. Time in Market | 6.92 days | 3.88 days | 9.96 days |
| **Profit per Month** | **3,623** | **1,583** | **2,047** |
| **Max. Time to Recover** | **58 days** | **56 days** | **92 days** |

*Table 3: Trading strategy statistics for Kalman filter model 4*

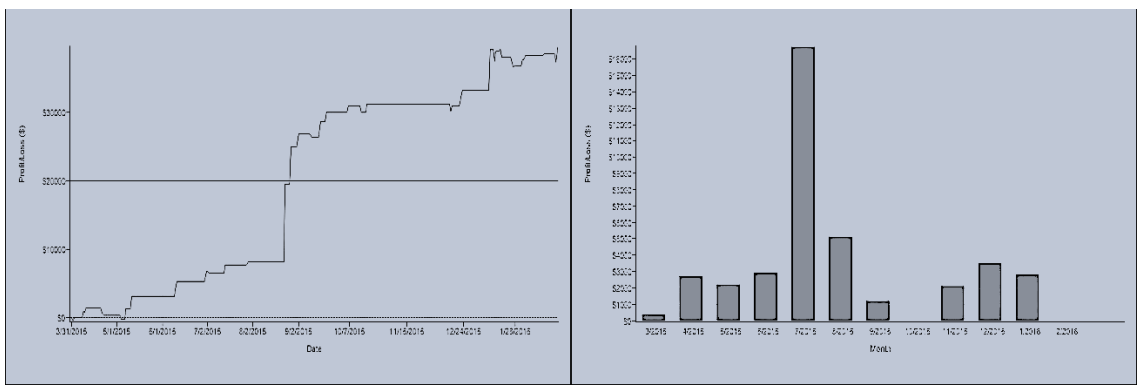

*Figure 5: Cumulative profit and monthly PnL distribution for KF model 4*

| Kalman Filer Model 3 | | | |
|---|---:|---:|---:|
| Field | All | Long | Short |
| **Net Profit (A+B)** | **29,022** | **12,009** | **17,013** |
| Gross Profit (A) | 47,548 | 24,403 | 23,145 |
| Gross Loss (B) | (18,526) | (12,394) | (6,132) |
| Total Commission | 228 | 116 | 112 |
| Drawdown | (3,820) | (3,366) | (1,579) |
| Sharpe Ratio | 1.22 | 0.38 | 1.65 |
| Profit Factor (A/B) | 2.57 | 1.97 | 3.77 |
| | | | |
| **Number of Trades** | **57** | **29** | **28** |
| Winning Trades | 35 | 18 | 17 |
| **Average Trade Profit** | **509** | **414** | **608** |
| Average Winning Trade | 1,359 | 1,356 | 1,361 |
| Largest Winning Trade | 5,234 | 5,234 | 4,271 |
| Max. conseq. Winners | 14 | 7 | 7 |
| | | | |
| Losing Trades | 22 | 11 | 11 |
| Average Losing Trade | (842) | (1,127) | (557) |
| Largest Losing Trade | (2,879) | (2,879) | (1,579) |
| Max. conseq. Losers | 5 | 3 | 3 |
| | | | |
| Ratio avg. Win / avg. Loss | 1.61 | 1.20 | 2.44 |
| Winning/Total | 0.61 | 0.62 | 0.61 |
| | | | |
| Avg. Time in Market | 5.82 days | 6.45 days | 5.18 days |
| **Profit per Month** | **2,658** | **1,100** | **1,860** |
| **Max. Time to Recover** | **53 days** | **64 days** | **71 days** |

*Table 4: Trading strategy statistics for Kalman filter model 3*

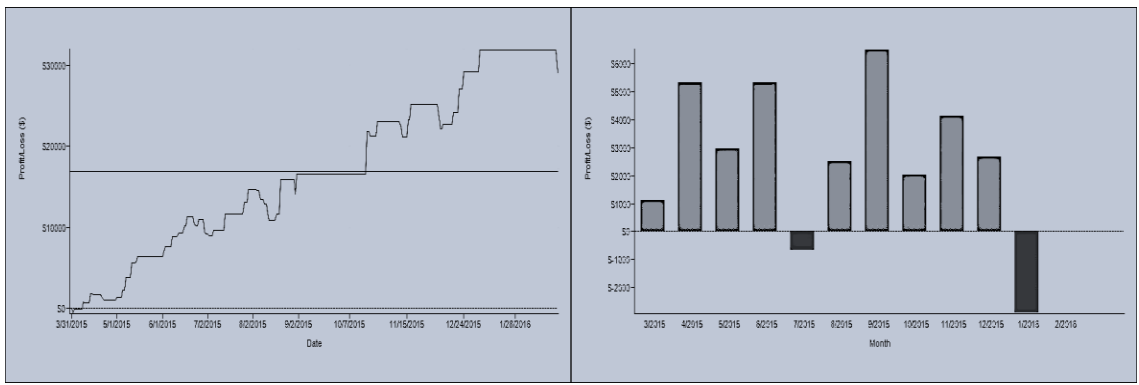

*Figure 6: Cumulative profit and monthly PnL distribution for KF model 3*

| Kalman Filer Model 2 | | | |
|---|---:|---:|---:|
| Field | All | Long | Short |
| **Total Net Profit** | **22,380** | **8,692** | **13,688** |
| Gross Profit | 40,747 | 19,611 | 21,136 |
| Gross Loss | (18,367) | (10,919) | (7,448) |
| Commission | 220 | 108 | 112 |
| Drawdown | (7,348) | (6,628) | (2,283) |
| Sharpe Ratio | 0.76 | 0.32 | 0.55 |
| Profit Factor | 2.22 | 1.80 | 2.84 |
| | | | |
| **Number of Trades** | **55** | **27** | **28** |
| Winning Trades | 32 | 16 | 16 |
| **Average Trade Profit** | **407** | **322** | **489** |
| Average Winning Trade | 1,273 | 1,226 | 1,321 |
| Largest Winning Trade | 6,521 | 3,221 | 6,521 |
| Max. conseq. Winners | 4 | 6 | 3 |
| | | | |
| Losing Trades | 23 | 11 | 12 |
| Average Losing Trade | (799) | (993) | (621) |
| Largest Losing Trade | (3,679) | (3,679) | (1,717) |
| Max. conseq. Losers | 4 | 5 | 2 |
| | | | |
| Ratio avg. Win / avg. Loss | 1.59 | 1.23 | 2.13 |
| Winning/Total | 0.58 | 0.59 | 0.57 |
| | | | |
| Avg. Time in Market | 6.04 days | 6.41 days | 5.68 days |
| **Profit per Month** | **2,050** | **801** | **1,254** |
| **Max. Time to Recover** | **132 days** | **146 days** | **70 days** |

*Table 5: Trading strategy statistics for Kalman filter model 2*

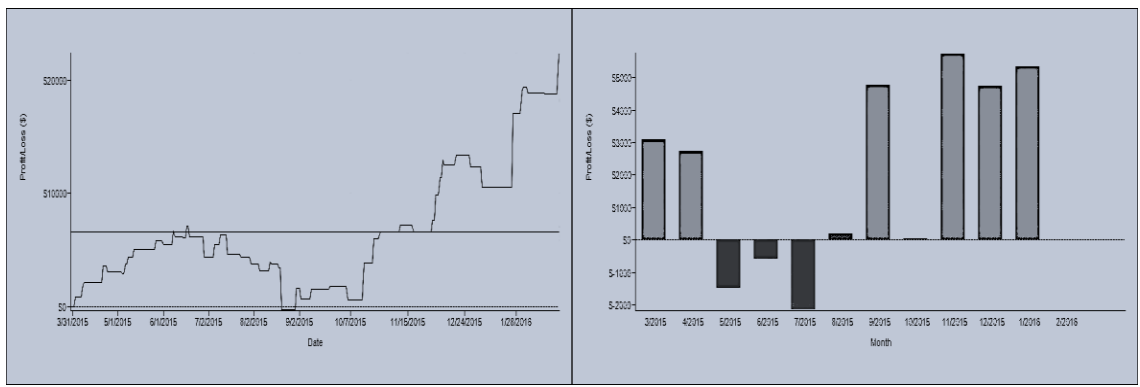

*Figure 7: Cumulative profit and monthly PnL distribution for KF model 2*

| Kalman Filer Model 1 | | | |
|---|---:|---:|---:|
| Field | All | Long | Short |
| **Total Net Profit** | **18,755** | **6,880** | **11,876** |
| Gross Profit | 39,151 | 18,861 | 20,290 |
| Gross Loss | (20,396) | (11,982) | (8,415) |
| Commission | 220 | 108 | 112 |
| Drawdown | (7,348) | (6,628) | (2,283) |
| Sharpe Ratio | 0.72 | 0.26 | 0.48 |
| Profit Factor | 1.92 | 1.57 | 2.41 |
| | | | |
| **Number of Trades** | **55** | **27** | **28** |
| Winning Trades | 31 | 16 | 15 |
| **Average Trade Profit** | **341** | **255** | **424** |
| Average Winning Trade | 1,263 | 1,179 | 1,353 |
| Largest Winning Trade | 6,521 | 3,221 | 6,521 |
| Max. conseq. Winners | 4 | 6 | 2 |
| | | | |
| Losing Trades | 24 | 11 | 13 |
| Average Losing Trade | (850) | (1,089) | (647) |
| Largest Losing Trade | (3,679) | (3,679) | (1,717) |
| Max. conseq. Losers | 4 | 5 | 2 |
| | | | |
| Ratio avg. Win / avg. Loss | 1.49 | 1.08 | 2.09 |
| Winning/Total | 0.56 | 0.59 | 0.54 |
| | | | |
| Avg. Time in Market | 6.04 days | 6.45 days | 5.64 days |
| Profit per Month | 1,718 | 634 | 1,088 |
| **Max. Time to Recover** | **132 days** | **146 days** | **71 days** |

*Table 6: Trading strategy statistics for Kalman filter model 1*

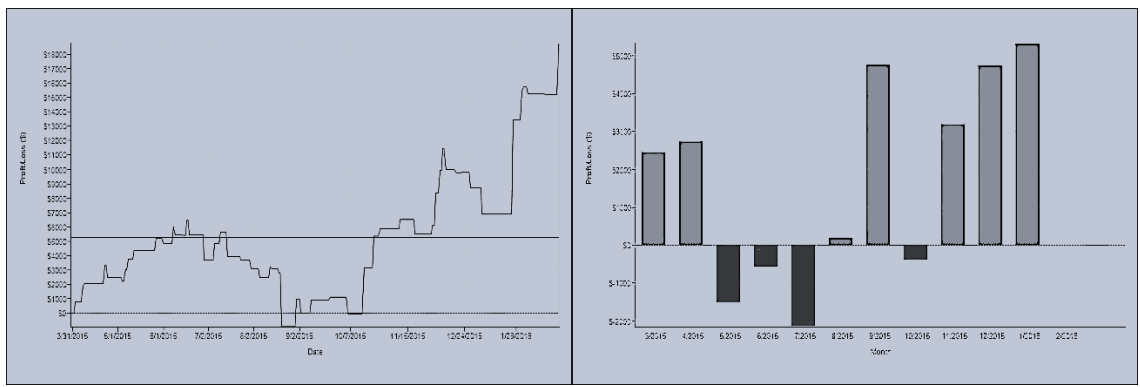

*Figure 8: Cumulative profit and monthly PnL distribution for KF model 1*

| Trade | Direction | Entry date | Entry price | Exit date | Exit price | Profit | PnL | Commission | Days in position |
|---|---|---|---|---|---|---|---|---|---|
| 1 | Long | Mar-31-15 | 2,043 | Apr-01-15 | 2033.5 | (454.0) | (454.0) | 4 | 2 |
| 2 | Short | Apr-01-15 | 2,034 | Apr-02-15 | 2026 | 371.0 | (83.0) | 4 | 2 |
| 3 | Long | Apr-02-15 | 2,026 | Apr-08-15 | 2045.25 | 958.5 | 875.5 | 4 | 4 |
| 4 | Short | Apr-08-15 | 2,045 | Apr-09-15 | 2047.5 | (116.5) | 759.0 | 4 | 2 |
| 5 | Long | Apr-09-15 | 2,048 | Apr-10-15 | 2061.5 | 696.0 | 1,455.0 | 4 | 2 |
| 6 | Short | Apr-10-15 | 2,062 | Apr-21-15 | 2076.25 | (741.5) | 713.5 | 4 | 8 |
| 7 | Long | Apr-21-15 | 2,076 | Apr-22-15 | 2069.75 | (329.0) | 384.5 | 4 | 2 |
| 8 | Short | Apr-22-15 | 2,070 | May-04-15 | 2081.75 | (604.0) | (219.5) | 4 | 9 |
| 9 | Long | May-04-15 | 2,082 | May-05-15 | 2079.75 | (104.0) | (323.5) | 4 | 2 |
| 10 | Short | May-05-15 | 2,080 | May-07-15 | 2048.25 | 1,571.0 | 1,247.5 | 4 | 3 |
| 11 | Long | May-07-15 | 2,048 | May-11-15 | 2084.75 | 1,821.0 | 3,068.5 | 4 | 3 |
| 12 | Short | May-11-15 | 2,085 | Jun-10-15 | 2062.25 | 1,121.0 | 4,189.5 | 4 | 23 |
| 13 | Long | Jun-10-15 | 2,062 | Jun-11-15 | 2083.5 | 1,058.5 | 5,248.0 | 4 | 2 |
| 14 | Short | Jun-11-15 | 2,084 | Jul-01-15 | 2054.25 | 1,458.5 | 6,706.5 | 4 | 15 |
| 15 | Long | Jul-01-15 | 2,054 | Jul-03-15 | 2049.75 | (229.0) | 6,477.5 | 4 | 3 |
| 16 | Short | Jul-03-15 | 2,050 | Jul-10-15 | 2050.5 | (41.5) | 6,436.0 | 4 | 6 |
| 17 | Long | Jul-10-15 | 2,051 | Jul-14-15 | 2074.5 | 1,196.0 | 7,632.0 | 4 | 3 |
| 18 | Short | Jul-14-15 | 2,075 | Jul-29-15 | 2071 | 171.0 | 7,803.0 | 4 | 12 |
| 19 | Long | Jul-29-15 | 2,071 | Jul-30-15 | 2077.5 | 321.0 | 8,124.0 | 4 | 2 |
| 20 | Short | Jul-30-15 | 2,078 | Aug-24-15 | 1851.25 | 11,308.5 | 19,432.5 | 4 | 18 |
| 21 | Long | Aug-24-15 | 1,851 | Aug-28-15 | 1960 | 5,433.5 | 24,866.0 | 4 | 5 |
| 22 | Short | Aug-28-15 | 1,960 | Sep-02-15 | 1921.25 | 1,933.5 | 26,799.5 | 4 | 4 |
| 23 | Long | Sep-02-15 | 1,921 | Sep-10-15 | 1920.25 | (54.0) | 26,745.5 | 4 | 7 |
| 24 | Short | Sep-10-15 | 1,920 | Sep-11-15 | 1928.25 | (404.0) | 26,341.5 | 4 | 2 |
| 25 | Long | Sep-11-15 | 1,928 | Sep-17-15 | 1974.75 | 2,321.0 | 28,662.5 | 4 | 5 |
| 26 | Short | Sep-17-15 | 1,975 | Sep-21-15 | 1948.5 | 1,308.5 | 29,971.0 | 4 | 3 |
| 27 | Long | Sep-21-15 | 1,949 | Oct-06-15 | 1967.5 | 946.0 | 30,917.0 | 4 | 12 |
| 28 | Short | Oct-06-15 | 1,968 | Oct-15-15 | 1986.25 | (941.5) | 29,975.5 | 4 | 8 |
| 29 | Long | Oct-15-15 | 1,986 | Oct-19-15 | 2010.25 | 1,196.0 | 31,171.5 | 4 | 3 |
| 30 | Short | Oct-19-15 | 2,010 | Dec-15-15 | 2031 | (1,041.5) | 30,130.0 | 4 | 42 |
| 31 | Long | Dec-15-15 | 2,031 | Dec-16-15 | 2048.25 | 858.5 | 30,988.5 | 4 | 2 |
| 32 | Short | Dec-16-15 | 2,048 | Dec-22-15 | 2022.75 | 1,271.0 | 32,259.5 | 4 | 5 |
| 33 | Long | Dec-22-15 | 2,023 | Dec-23-15 | 2043 | 1,008.5 | 33,268.0 | 4 | 2 |
| 34 | Short | Dec-23-15 | 2,043 | Jan-11-16 | 1924.5 | 5,921.0 | 39,189.0 | 4 | 12 |
| 35 | Long | Jan-11-16 | 1,925 | Jan-14-16 | 1890 | (1,729.0) | 37,460.0 | 4 | 4 |
| 36 | Short | Jan-14-16 | 1,890 | Jan-15-16 | 1862 | 1,396.0 | 38,856.0 | 4 | 2 |
| 37 | Long | Jan-15-16 | 1,862 | Jan-18-16 | 1869.5 | 371.0 | 39,227.0 | 4 | 2 |
| 38 | Short | Jan-18-16 | 1,870 | Jan-19-16 | 1894 | (1,229.0) | 37,998.0 | 4 | 2 |
| 39 | Long | Jan-19-16 | 1,894 | Jan-26-16 | 1878.5 | (779.0) | 37,219.0 | 4 | 6 |
| 40 | Short | Jan-26-16 | 1,879 | Jan-27-16 | 1890.25 | (591.5) | 36,627.5 | 4 | 2 |
| 41 | Long | Jan-27-16 | 1,890 | Jan-28-16 | 1894 | 183.5 | 36,811.0 | 4 | 2 |
| 42 | Short | Jan-28-16 | 1,894 | Jan-29-16 | 1894.5 | (29.0) | 36,782.0 | 4 | 2 |
| 43 | Long | Jan-29-16 | 1,895 | Feb-02-16 | 1913.5 | 946.0 | 37,728.0 | 4 | 3 |
| 44 | Short | Feb-02-16 | 1,914 | Feb-04-16 | 1901.5 | 596.0 | 38,324.0 | 4 | 3 |
| 45 | Long | Feb-04-16 | 1,902 | Feb-17-16 | 1906 | 221.0 | 38,545.0 | 4 | 10 |
| 46 | Short | Feb-17-16 | 1,906 | Feb-25-16 | 1931.25 | (1,266.5) | 37,278.5 | 4 | 7 |
| 47 | Long | Feb-25-16 | 1,931 | Feb-26-16 | 1959.75 | 1,421.0 | 38,699.5 | 4 | 2 |
| 48 | Short | Feb-26-16 | 1,960 | Feb-26-16 | 1942.5 | 858.5 | 39,558.0 | 4 | 1 |

*Table 7: Trades list for Kalman model 4*

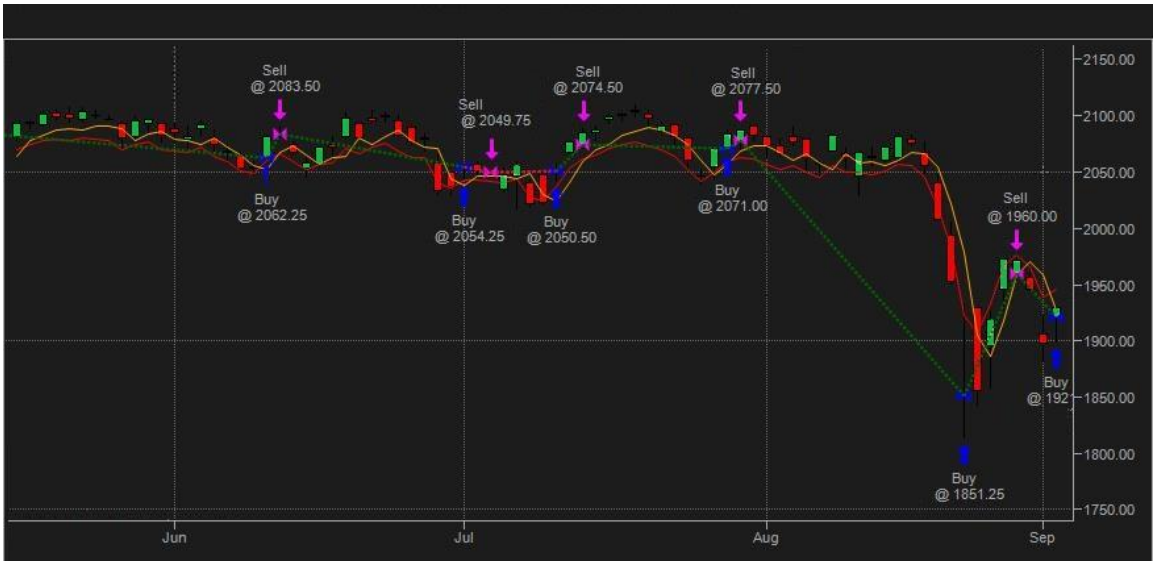

*Figure 9: Efficiency of Kalman filter model 4 to detect trends*

# 6 DISCUSSION

Parameters for the Kalman filter models are obtained by a general optimization. Hence they provide the best possible choice of parameters. Results presented here should be analyzed with this in mind.

We clearly see that model 1 and 2 provides similar results about 20 k of net profit for one year trading the E-mini contract. When adding the new feature of a short and long term model factor, we increase net profit to 29 k, which is substantial. We reduce maximum drawdown from -7,300 USD to -3,800 USD. This is a material gain. Model 4 performs even better as we generate an additional 10 k with net profit skyrocketing to 40 k USD with an further reduction of drawdown to -2,600 USD.

# CONCLUSION

In this paper, we empirically validate that Kalman filters with meaningful dynamics have predictive power. After reviewing moving averages and the general equation for its lag at order n with respect to the one at first order, we examine 4 Kalman filter models: the common one with speed and acceleration concepts, the traditional statistical one referred to as the local linear trend, a new model that splits price contribution between short and long term effect and a last one that encompasses all above with an additional term corresponding to the position of the price with regards to its extremums. We find empirically that model 4 performs far better than any other models. We also confirm that KF models have zero lag and capture price dynamic better than previous combinations of moving averages like DEMA or TEMA. We confirm on model 4 that oscillators and trend following indicators are a powerful combination that performs better than any single indicators.

# APPENDIX

**Proof A.1:**

We want to prove

$$LAG\ MA\odot^k(w_0, \ldots, w_N) = k\ LAG\ MA(w_0, \ldots, w_N) \quad (EQ.2.7)$$

For k=1, this result is obviously true. If the result is true at order $k-1$, let us show that it holds at order $k$. We want to compute the following lag

$$LAG\ MA\odot^k(w_0, \ldots, w_N) = \frac{\sum_{i_1=0,\ldots,i_k=0}^{N} w_{i_1} \ldots w_{i_k} T_{i_1+\cdots+i_k}}{\sum_{i_1=0,\ldots,i_k=0}^{N} w_{i_1} \ldots w_{i_k}} \quad (A.1.1)$$

Which can be reformulated as:

$$\frac{\sum_{i_1=0,\ldots,i_{k-1}=0}^{N} w_{i_1} \ldots w_{i_{k-1}} \sum_{i_k=0}^{N} w_{i_k}\left(T_{i_1+\cdots+i_{k-1}} + T_{i_k}\right)}{\sum_{i_1=0,\ldots,i_{k-1}=0}^{N} w_{i_1} \ldots w_{i_{k-1}} \sum_{i_k=0}^{N} w_{i_k}} \quad (A.1.2)$$

Or

$$\frac{\sum_{i_1=0,\ldots,i_{k-1}=0}^{N} w_{i_1} \ldots w_{i_{k-1}} T_{i_1+\cdots+i_{k-1}}}{\sum_{i_1=0,\ldots,i_{k-1}=0}^{N} w_{i_1} \ldots w_{i_{k-1}}} + \frac{\sum_{i_k=0}^{N} w_{i_k} T_{i_k}}{\sum_{i_k=0}^{N} w_{i_k}} \quad (A.1.3)$$

Or

$$= (k-1)\ LAG\ MA(w_0, \ldots, w_N) + LAG\ MA(w_0, \ldots, w_N) \quad (A.1.4)$$

Which proves the result!

**Proof A.2:**

If we denote by a and b the coefficient of the linear combination between the moving average and the second order recursive moving average, the resulting formula should be

$$aMA + bMA\odot^2 \quad (A.2.1)$$

With the constraints that first, the sum of the coefficients should be equal to 1, hence

$$a + b = 1 \quad (A.2.2)$$

Second, the lag of the resulting combination should be 0, hence

$$a + 2b = 0 \quad (A.2.3)$$

It is easy to solve this linear system and find

$$a = 2, \quad b = -1 \qquad (A.2.4)$$

Which is the result!

**Proof A.3:**

If we denote by a, b and c the coefficient of the linear combination between the moving average, the second and the third order recursive moving average, the resulting formula should be

$$aMA + bMA\odot^2 + cbMA\odot^3 \qquad (A.3.1)$$

With the following constraints. First, the sum of the coefficients should be equal to 1, hence

$$a + b + c = 1 \qquad (A.3.2)$$

Second, the lag of the resulting combination should be 0, hence

$$a + 2b + 3c = 0 \qquad (A.3.3)$$

Third, we impose that the coefficient for the third order recursive moving average is equal to 1,

$$c = 1 \qquad (A.3.4)$$

It is easy to solve this linear system and find

$$a = 3, \quad b = -3, \quad c = 1 \qquad (A.3.5)$$

Which is the result!